# SSD FORENSIC: EVIDENCE GENERATION AND FORENSIC RESEARCH ON SOLID STATE DRIVES USING TRIM ANALYSIS

Hassan Jalil Hadi, Irshad ullah khan, Sheetal Harris
Email: hjalil.buic@bahria.edu.pk, irrshad@gmail.com,

# Abstract

Traditional hard drives consisting of spinning magnetic media platters are becoming things of the past as with the emergence of the latest digital technologies and electronic equipment, the demand for faster, lighter, and more reliable alternate storage solutions is imperative. To attain these requirements, flash storage technologies like Solid State Drive (SSD) has overtaken traditional hard disk drives. In a forensic analysis of flash storage devices, forensic investigators are facing severe challenges for the reason that the sovereign behavior of solid-state storage media does not look favorable compared to traditional storage media devices. Wear Leveling, a fundamental mechanism in Solid State Drive (SSD), plays a severe challenge that most often destroys forensic evidence in many cases. It makes it complicated for forensic investigators to recover the necessary evidence. Persistence of deleted data in flash storage media depends on various factors like the Garbage Collection process, TRIM command, flash media type, manufacturer, capacity, file system, type of file saved, and the Operating System, etc. In view of this, extensive experiments conducted to identify the probability of data recovery and carving. Analyzed effects of Wear Leveling and Garbage Collection processes in Solid State Drive (SSD) of different manufacturers, having the same storage capacities and with a different type of files utilized. In conclusion, experimental findings established the fact that Wear Leveling in solid-state media can obfuscate digital evidence, and a conventional assumption regarding the behavior of storage media is no more valid. Moreover, data persistency also depends on the manufacturers, time-lapse of forensic analysis after data deletion, type of files, and size of files stored in Solid State Drives (SSD).

## 1.1 Introduction

In the last few years, digital storage devices have been revolutionized, and its storage capacities, performance, and reliability improved significantly. Traditional storage drives equipped with rotational magnetic platters are now decreasing by utilization in the latest digital equipment. Flash-based memory has been emerged for a decade and gaining its popularity with every passing day due to its speed and performance over traditional media drives.

Conventional storage drives with magnetic storage components have widely known forensic properties. Computational activities which depend on some storage access leave traces, which can be identified later in the case of any forensic investigations. As Solid State Drive (SSD) is a recent technology, been used in today's most electronic applications contains various advantages like read/write speed, performance, noise-less, and endurance against shocks and temperature. However, Solid State Drive (SSD) has certain inherent limitations such as short life of flash memory cells concerning data writes, and it is necessary to erase data from the drive before any further write/re-write on the same blocks [1].

In flash storage devices, numbers of write cycles on storage cells are limited, so different flash management techniques are used by the manufacturers to overcome premature flash wears off issues. One basic approach used in almost all flash-based storage devices is Wear Leveling. In this technique, the controller of Solid State Drive (SSD) tries to equalize the distribution of write/erase process on each available block [2]. In flash storage, erasure must perform earlier before any data writes. To figure out the issue of deleting pages before any data writes, a TRIM command feature used in all Solid State Drives (SSDs). It notifies the SSD controller that specific data blocks aren't being used (deleted) and is safe to wipe out internally. Modern Operating Systems, file systems, hardware interfaces like SATA/e-SATA, SCSI, and different SSDs all support TRIM command functionality [3].

Garbage Collection is the process that practically wipes unused pages in Solid State Drive. TRIM and Garbage Collection are the two sources that could destroy the evidence in SSD [4]. It is evident from existing researches and their analysis about the Solid State Drive (SSD) that it destroys forensic information, and the chances of recovery are almost unachievable [5].

Several types of research on Solid State Drive (SSD) in view of forensics investigation has carried out with legal concerns and difficulty in evidence recoveries, but no comprehensive way out is effectuated. Bell et al. [5] indicated that existing methods of evidence acquisition and tools are not anymore valid in Solid State Drives, and they're incapable of serving in lots of situations.

In digital forensics, traditional tools and mechanisms used for the acquisition of evidence from conventional hard drives (HDD), as they are working just the same from the last decades. However, these tools and methods cannot apply to Solid State Drives (SSD) because of some advance and inherent features like Wear Leveling, TRIM, and Garbage Collection process.

A research problem is to find an ultimate cause and critical factor that are inducing challenges to forensic detectives in seeking digital crime facts from Solid State Drives and analyze the results acquired from these different types of storage media devices.

# Literature Review
## 2.2    Related Work / Literature Survey

During the last few years, Solid State Drive (SSD) has proven significant success by its popularity of gaining and utilization in today's electronic devices due to its essential features of reliability and performance. Traditional magnetic media hard drives are phasing out gradually, and Solid State Drives (SSD) is filling the space of storage needs.

The literature reviewed in this chapter will provide brief research work conducted by most experts and researchers related to the forensics of Solid State Drive (SSD). Their area of work, experiments, tools, and methods used during their research experiments, and their findings well explained. It will give insights into different experimental works, addressed techniques and technologies of digital forensics.

Ashar et al. [6] presented Wear Leveling analysis in different solid state devices. The study conducted with four types of storage devices, i.e., USB flash, SD and MicroSD cards, and Solid State Drive (SSD). To observe Wear Leveling behavior in different types of files and Operating Systems, a single image file, including some media files selected to store on all devices. After delete and wiped out all cards/disks with professional erasing software, analyzed the result of data recovery and carving with the TRIM command enable and disable. It concluded that no single trace of previous records retrieved, and TRIM plays a crucial role in destroying digital evidence in flash-based storage devices.

Gubanov et al. [7] published an article regarding Solid State Drive (SSD) forensics explaining about self-corrosion phenomena in flash storage devices. He also explained the TRIM and Garbage Collection processes in detail. In his research work, he proposed a complicated solution of a chip off the controller of SSD to avoid executing the internal Garbage Collection process. Not only it requires special expert skills and tools, but it does not work in the case of encrypted data drives. Their research also explored the effects of the TRIM command in solid-state media. Later on, experiments on the behavior of eMMC chips in comparison to SSD conducted. It concluded that SSD forensics remains different from conventional hard drives. It destroys data after delete or format with autonomous behavior of background Garbage Collection process in all Solid State Drives (SSDs).

Binaya et al. [3] presented an analysis of Solid State Drives (SSDs). The study conducted on a single SSD drive in a working workstation by a TRIM enabled and disabled option. Multiple types of Operating Systems used with write blocking device to identify the difference in the analysis of forensic results. In their research work, it concluded that by enabling/disabling TRIM function in multiple Operating Systems, the behavior of data I/O changed, and the performance of data write decreased. If TRIM is enabled and in data delete/format, then chances of recovery in SSD is almost impossible. It makes it complicated for an investigator to recover the necessary evidence.

Alastair et al. [8] presented data retention analysis on Solid State Drives (SSD) with TRIM enabled file systems. The experiment conducted on SSD with three different file systems of Microsoft Windows, MAC, and Linux, namely NTFS, HFS+, and Ext4, respectively. It ran under some specific conditions like manual passing of TRIM command enable/disable, low and high drive usage, and in idle workload situations. In conclusion, the result of experiments and research work showed that TRIM plays an essential role in SSDs Wear Leveling functionality, and the evidence in the case of TRIM enable is almost unrecoverable. The behavior of NTFS and HFS+ for the TRIM command is nearly the same, whereas the Ext4 file system in Linux Operating System behaves differently. It executed commands in batches; therefore, a chance of data retrieval after deletion within a short time is somehow possible.

Bednar et al. [9] published a paper presenting some challenges in view of Solid State Drive (SSD) forensics. The work elaborated on principles and unexpected difficulties that forensic investigators faced during SSD analysis. The conclusion is to adopt conventional tools and guidelines in conducting traditional hard drive investigations that are not viable in the case of Solid State Drives (SSD) because both of them contain fundamental differences in a storage mechanism. Moreover, write blockers in case of SSDs cannot fulfill its purpose. Even by using write blockers with SSDs, background processes like Garbage Collection couldn't control, and the required evidence will not be recoverable.

King et al. [10] presented an analysis of their detailed study in 2011. Different types of fifteen SSDs and one HDD drive used in the research to check the admissibility of data missing and recovery. It concluded in the study that with the TRIM enabled option, no data was able to recover, but recovery was reasonably possible when TRIM feature manually disabled within the Operating System. Recovery of data differed from vendor to vendor in all SSD drives. The author concluded that traditional data recovery techniques and methods are not possible in the case when the TRIM feature remains enabled for SSDs.

Freeman et al. [11] published his experimental work and their analysis on Solid State Drive (SSD). In his demonstration and conclusion, it's impossible to recover any data if SSD is wiped securely with any professional data erasing software. Some files can carve, but those are not useful for evidence. The primary goal of this research was to check the efficiency of safe data removal in Solid State Drives.

Zubair et al. [21] presented an analysis of their work on Solid State Drive (SSD) connected with different data interfaces, i.e., USB, primary SATA, and secondary SATA interfaces. In his research, issues of evidence destruction in solid-state media well-investigated and reported his experimental findings. In conclusion of these experiments, it is clear that behaviors of SSDs are different when connected to USB, primary, and secondary SATA interfaces. TRIM command only works when SSD connected to the primary SATA interface. In case SSD is connected in USB or secondary SATA interfaces, the functionality of the TRIM command does not work. Therefore, the background Garbage Collection process will not execute, and deleted data is recoverable.

Vieyra et al. [12] published his experimental work regarding the analysis of Solid State Drive (SSD). Purpose of research work involved in finding the effects of background processes and guidance of its function during a forensic investigation. Points like extraction of artifacts, under different conditions of data sizes, power impacts, etc. were in consideration. The author concluded that even SSD works differently, but forensic should stay practically the same. NAND flash may hold deleted data, but an Operating System is unable to show it. Currently, it is a general practice by forensic investigators that always collect the image of data evidence with write blockers to maintain its integrity, but in the future, it may be some chip off mechanisms that would be able to access deleted data. Currently, the chip off method is not in practice, and drawbacks explained in the articles of [7].

Marupudi et al. [13] conducted research work on traditional storage drive (HDD) against Solid State Drive (SSD). In his work, same type of files in both HDD and SDD drives used. Formatted both drives at the same time intervals and stored the same files therein. Image the drives with FTK Imager and analyzed the behavior in-depth, it concluded that even on both drives, the treatment was the same, but after studying the result on both drives was entirely different. Based on the results obtained, it is clear that SSD with TRIM, Wear Leveling, and Garbage Collection processes destroyed evidence, and same is not in the case of traditional hard drives.

Kambalapalli et al. [14] presented an analysis of their work on HDD and SDD with different forensic tools, their benefits, and drawbacks. For this research, the same single file was created and used in both storage devices. HD Shredder software used to format the disk and image taken for analysis with different forensic tools. Based on these experiments and after research, it proved that SSDs destroyed evidence or lost critical data, which may create difficulties for the forensic investigators by using existing tools and techniques.

B. Bell et al. [7] presented their research work with Solid State Drive (SSD) and traditional storage devices. Research work conducted on single SSD and HDD with filling textual data. Quick format the drives and immediately powered off the system. At power-on, the process of garbage collection started instantly, and within a few minutes, data from the drive was unrecoverable. Purging of evidence took place even drives attached with physical write blockers. In their research findings, it concluded that SSD could destroy evidence under its preference if instructions passed to the SSD controller, and there is no method to stop it even by powering off the whole system. The process will immediately start again as the drive gets power, and it takes very few moments to complete.

John Fulton [15] submitted his project work explaining in detail the issues and limitations of solid-state forensics. The focus of the research was to sort out problems that exist for forensic investigators and manufacturers of different flash storage devices. His recommendation was to provide detailed information about the algorithms and methods used in the clearing process of deleted files. There could be one protocol that examines all the details of the clearing process and the possibility of provision physical on/off button in Solid State Drives.

Josue Ferreira [16] submitted an article with detailed work about the acquisition of forensic evidence from Solid State Drive (SSD) by using Open-source tools. His work proposed a method to perform forensic data acquisition from SSDs while preventing the TRIM and garbage collection process during the investigation and evidence collection. By disabling automount feature or use forensic live CDs, connecting with write blocker equipment, and with some more integrity check methods, a forensic investigator can prevent from any alteration in binary sequence (bit-stream) copy of the evidence. The author also concluded that the hash value of binary sequence copy, i.e., bit-stream, differs when a drive connected with different interfaces of a forensic workstation.

## 2.3   Comparison of Existing Work and Techniques

In digital forensics, lots of research work, including detailed experiments on conventional storage drives performed. Their tools, methods, and guidelines well understood among the community of forensic practitioners. Solid State Drive (SSD) with flash cells is a new technology; therefore, there exist some limitations of tools and practices that forensic investigators are not well informed. Here is a brief table comparing existing work in SSDs:

Table 2.1: Comparison of existing work

| Sr. | Paper / Article | Forensic Analysis | | | | | | | |
|---|---|---|---|---|---|---|---|---|---|
| | | SSD | Same Media | Same Capacity | Diff Files | Diff Size | Diff O.S | Idle Time | Recovery |
| 01 | Ashar et. al | Yes | No | No | Yes | Yes | Yes | > 1 Day | No |
| 02 | Gubanov et. al | Yes | No | No | No | No | - | - | No |
| 03 | Binaya et. al | Yes | No | No | Yes | No | No | - | Yes (TRIM off) |
| 04 | Alastair et.al | Yes | No | No | No | No | Yes | - | Yes (TRIM off) |
| 05 | Bednar et. al | Yes | No | No | No | No | No | - | No |
| 06 | King et. al | Yes | No | No | Yes | No | No | - | No |
| 07 | Freeman et. al | Yes | No | No | No | No | No | - | No |
| 08 | Zubair et. al | Yes | No | No | Yes | No | Yes | - | No |
| 09 | Vieyra et. al | Yes | No | No | Yes | Yes | No | - | No |
| 10 | Marupudi et. al | Yes | No | No | No | No | No | - | No |
| 11 | Kambalapalli et. al | Yes | No | No | No | No | No | - | No |

| 12 | B. Bell et. al | Yes | No | No | No | No | No | < 1 min | No |
| 13 | John Fulton | Yes | No | No | Yes | No | No | - | No |
| 14 | Josue Ferreira | Yes | No | No | Yes | No | No | - | No |
| | **Proposed Work** | **Yes** | **Yes** | **Yes** | **Yes** | **Yes** | **Yes** | **Min/Hr(s)** | **Uncertain** |

## 2.4 Limitations of Existing Work

Several studies have conducted for Solid State Drives (SSD) in different aspects of forensic research and the challenges faced by forensic experts. Various tools, techniques, and practices selected to explore the behavior of SSDs and to find the challenges faced by forensic investigators. Table 2.1 provides a detailed comparison of existing work by reviewing the literature, research work, and their conclusions. In the light of existing work in SSD forensics, following are the limitations of work and areas which have not touched so far:

- No research conducted so far on forensic behavior of different SSD manufacturers with the same drive capacities.
- No research conducted so far to find the file size and type affect the evidence recovery in SSDs from different manufacturers.
- The effect of evidence recovery by formatting and data deletion of different SSDs has not explored.
- The time difference between data deletion and image acquisition has not explored to examine the effect of evidence recovery results.

To address the limitations of existing work in SSD forensic, detailed research work is being conducted with the latest software, and guidelines precisely will cover identified deficiencies. It would be appropriate to develop awareness and knowledge of best practices on SSD forensics.

# Methodology

## 3.1 Design of Work

In this section, the discussion will be regarding research methodology and design of work coupled with requirements to perform forensic analysis of Solid State Drives (SSDs). We will discuss key concepts of flash technologies like the TRIM command feature, its scope in the Operating System, background Garbage Collection process, and how it plays a vital role in destroying evidence automatically after data delete or format. Besides, the discussion will be on research's workflow as well as software and hardware requirements.

## 3.2 Requirements

Following are a few requirements which need to fulfill before the start of research:

### 3.2.1 Hardware

- A desktop computer or a laptop with SATA optical disk drive installed ⬜ SSDs of different manufacturers with the same technical specifications
- HDD Caddy case of SATA interface
- An external data storage device for disk images backup (optional)

### 3.2.2 Software

- TRIM command supported Operating System
- Software for complete data erasure of Solid State Drives
- Software for disk image acquisition and forensic analysis
- Software to check data integrity after recoveries
- Software for detailed analysis of hardware specifications
- Software for a bunch of data files creator

A laptop equipped with an optical disk drive is sufficient to carry out all required technical tasks of the project. A laptop with an option of optical disk drive contains an internal SATA interface, can be used to connect SSD by using an external accessory, i.e., an HDD caddy case. It used to connect an extra storage drive with the laptop in place of an already installed optical disk drive. In this research, all Solid State Drives (SSDs) were used in the same caddy drive to connect and mount with the Operating System.

TRIM, an ATA command feature, informs the SSD controller that a particular data is deleted from the end-user and is no more required. Afterward, the Garbage Collection process executes and wipe out entirely that unused data. The TRIM command feature is only applicable and runs if SSD drives connected with SATA or SCSI interfaces. It does not work in USB, RAID, or NAS mode; therefore, not supported to carry out the TRIM process in these modes of Solid State Drives [17]. To execute the TRIM command and Garbage Collection process properly, file systems of drives minimum required with NTFS in the case of Microsoft Windows, HFS+ in MAC OS, and EXT4 file system in Linux Operating Systems .

The following hardware and software used to carry out overall research work. Detailed technical specifications are as below:

i) **Forensic Workstation**

Lenovo ThinkPad T420s, 8GB RAM, 320 GB HDD

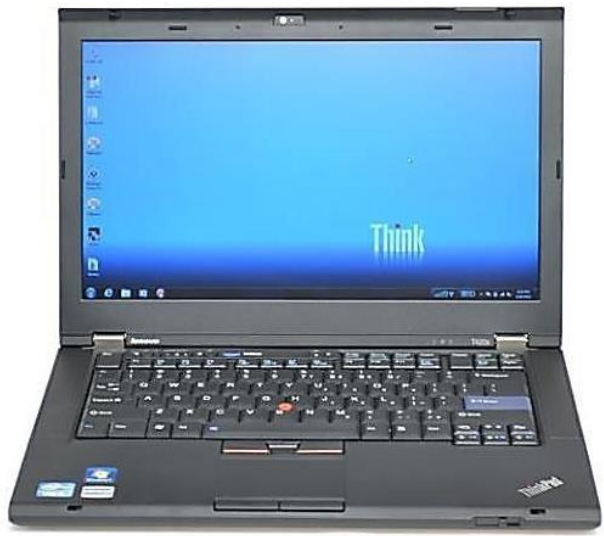

Figure 3.1: Forensic Workstation

ii) **Solid State Drive (SSD) 1**

Micron P400e, 2.5" 64 GB SSD, SATA 3 (6 Gbps)

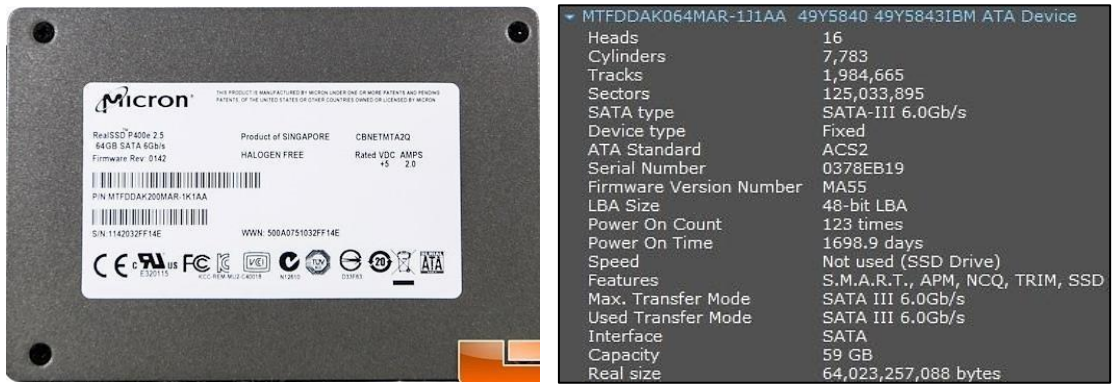

Figure 3.2:  Solid State Drive 1

iii)     **Solid State Drive (SSD) 2**

SanDisk SDSSDP, 2.5" 64 GB, SATA 3 (6 Gbps)

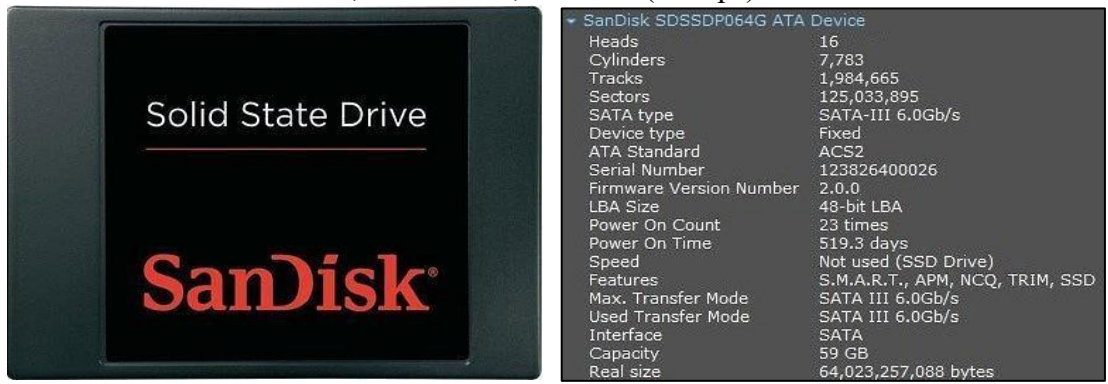

Figure 3.3:  Solid State Drive 2

iv)     **Solid State Drive (SSD) 3**

Transcend SSD370S, 2.5" 64 GB, SATA 3 (6 Gbps)

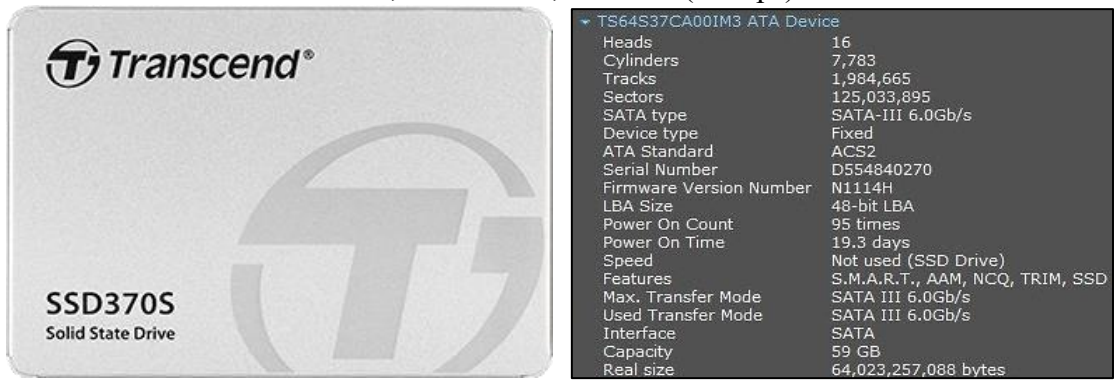

Figure 3.4:  Solid State Drive 3

v)     **HDD Caddy Case**

Chipal Universal SATA, 9.5mm 2nd HDD Caddy

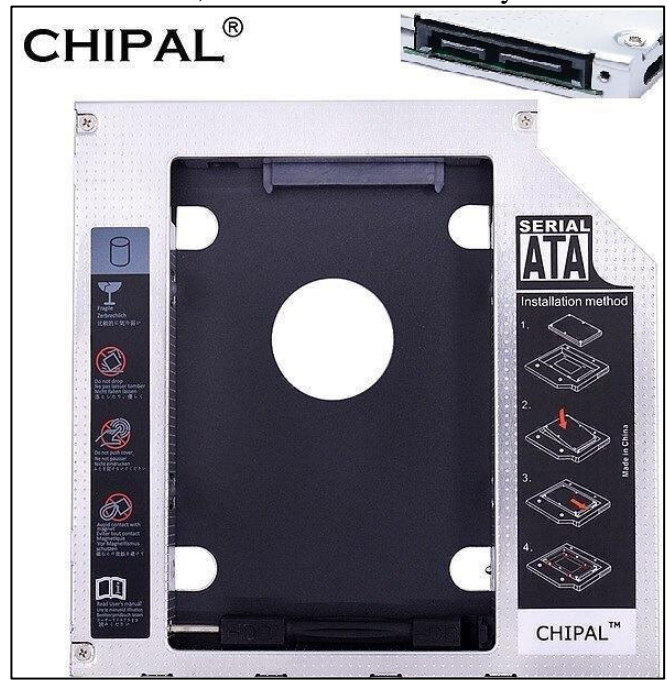

Figure 3.5: HDD Caddy Case

vi) **Operating System**

Microsoft Windows 7 Professional SP1, x64 bits

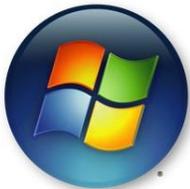

Figure 3.6: Operating System

vii) **FTK Imager (v 4.2.1.4)**

Software for disk image acquisition and forensic analysis

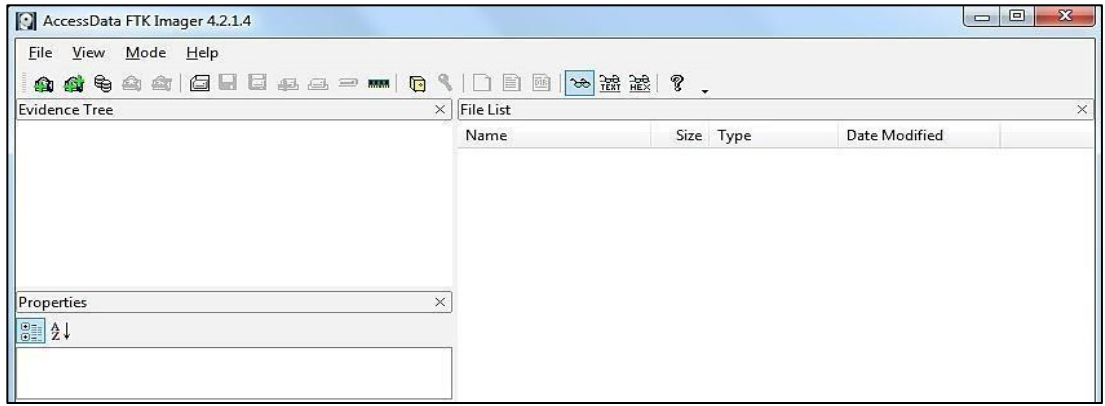

Figure 3.7: FTK Imager

viii) **Checksum Compare (v 1.42)**

Software for data integrity checks after retrieval

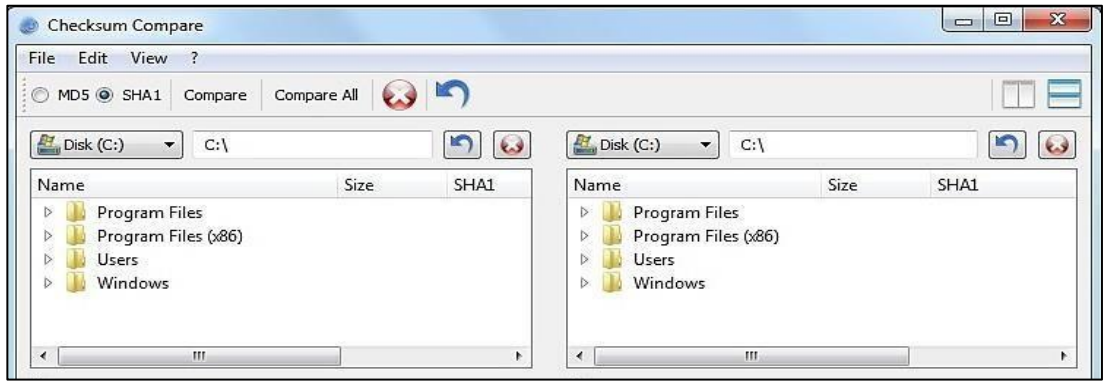

Figure 3.8: Checksum Compare

ix) **Eraser (v 6.2.0.2982)**

Software for complete data erasure from Solid State Drives

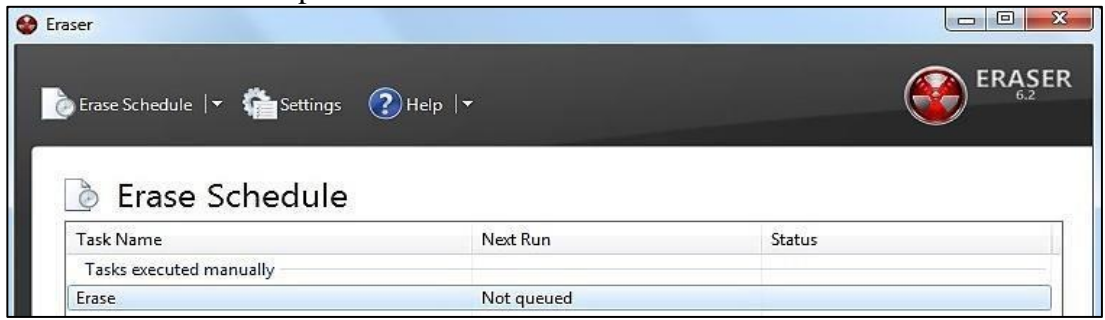

Figure 3.9: Eraser

x) **Speccy Professional (v 1.29.714) Portable**

Software for detailed analysis of hardware specifications

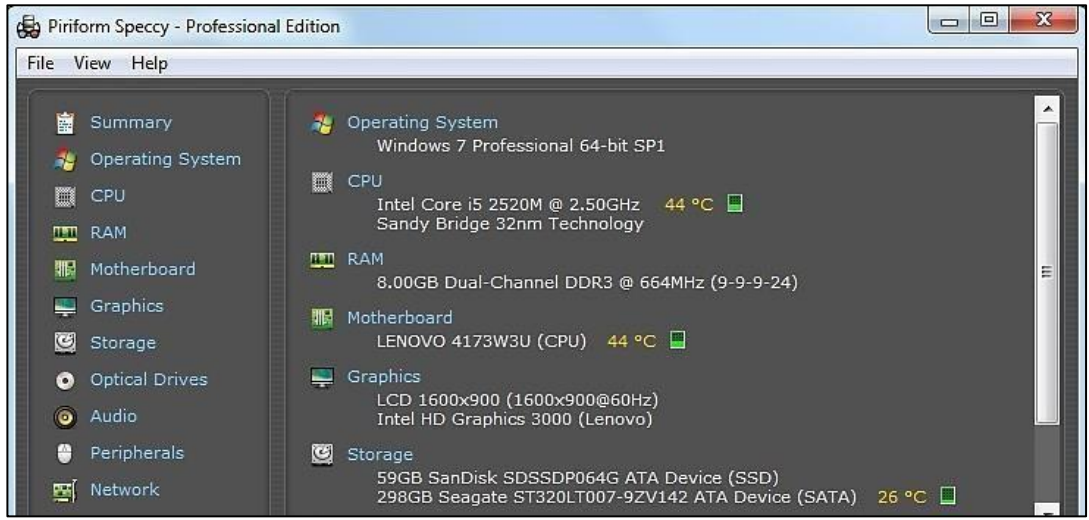

Figure 3.10: Speccy Professional

xi) **File Generator (v 3.6.0)**
Software for a bunch of data files creator

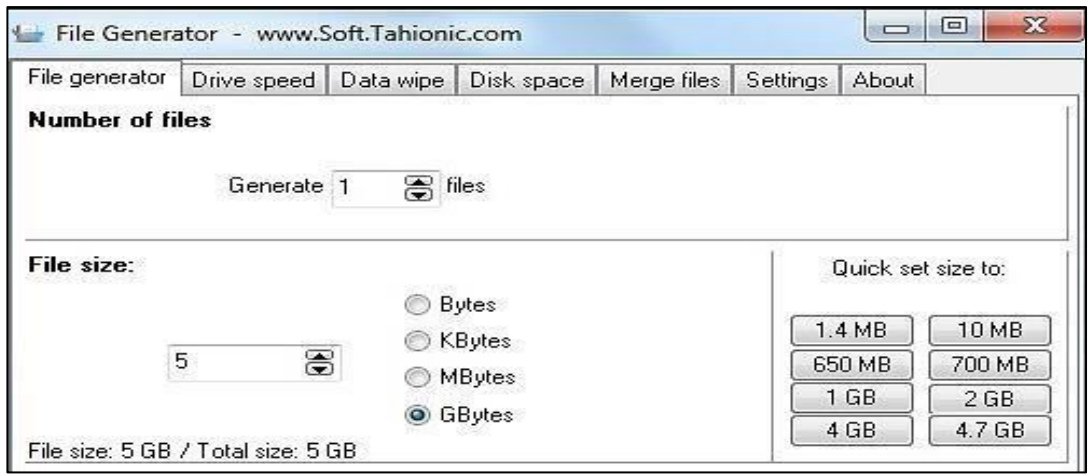

Figure 3.11: File Generator

### 3.2.3 Research Data Collection

For this research, many types of files and sizes required to observe the behavior of Wear Leveling in different SSD manufacturers having the same storage and technical specifications. In all test scenarios, a different type of files and sizes used which defined in the following four (04) data groups:

### 3.2.3.1 Data Set 1

This group contains the following type of files, sizes, and quantities:

Table 3.1: Data Set 1

| Sr # | File Type | File Size | Quantity | |
|---|---|---|---|---|
| 1 | **Image files** (jpeg, bmp, etc.) | 1 KB to 100 MB | 10 | Image file 01 — 60 KB<br>Image file 02 — 77 KB<br>Image file 03 — 289 KB<br>Image file 04 — 435 KB<br>Image file 05 — 862 KB<br>Image file 06 — 3,063 KB<br>Image file 07 — 11,282 KB<br>Image file 08 — 33,700 KB<br>Image file 09 — 40,066 KB<br>Image file 10 — 85,707 KB |
| 2 | **Text files** (txt, doc, xls, pdf, etc.) | 1 KB to 100 MB | 10 | Text file 01 — 37 KB<br>Text file 02 — 73 KB<br>Text file 03 — 524 KB<br>Text file 04 — 1,033 KB<br>Text file 05 — 4,716 KB<br>Text file 06 — 7,018 KB<br>Text file 07 — 17,445 KB<br>Text file 08 — 31,667 KB<br>Text file 09 — 45,395 KB<br>Text file 10 — 52,466 KB |
| 3 | **Media files** (mp3, mp4, m4v, etc.) | 1 KB to 100 MB | 10 | Media file 01 — 15 KB<br>Media file 02 — 1,268 KB<br>Media file 03 — 3,589 KB<br>Media file 04 — 14,153 KB<br>Media file 05 — 20,775 KB<br>Media file 06 — 32,537 KB<br>Media file 07 — 40,530 KB<br>Media file 08 — 50,879 KB<br>Media file 09 — 64,947 KB<br>Media file 10 — 80,632 KB |

| 4 | **Archives** (zip, rar, 7z, etc.) | 1 KB to 100 MB | 10 | Archive 01   13 KB<br>Archive 02   153 KB<br>Archive 03   3,518 KB<br>Archive 04   14,091 KB<br>Archive 05   20,681 KB<br>Archive 06   32,504 KB<br>Archive 07   40,529 KB<br>Archive 08   50,786 KB<br>Archive 09   64,927 KB<br>Archive 10   80,632 KB |
|---|---|---|---|---|
| 5 | **Disk images** (iso, bin, vcd, etc.) | 1 KB to 100 MB | 10 | Disk image 01   326 KB<br>Disk image 02   528 KB<br>Disk image 03   3,616 KB<br>Disk image 04   13,760 KB<br>Disk image 05   28,112 KB<br>Disk image 06   29,662 KB<br>Disk image 07   39,160 KB<br>Disk image 08   56,252 KB<br>Disk image 09   85,696 KB<br>Disk image 10   97,866 KB |
| 6 | **Data files** (dat, pst, sql, log, etc.) | 1 KB to 100 MB | 10 | Data file 01   95 KB<br>Data file 02   265 KB<br>Data file 03   885 KB<br>Data file 04   1,928 KB<br>Data file 05   2,249 KB<br>Data file 06   3,515 KB<br>Data file 07   10,185 KB<br>Data file 08   30,629 KB<br>Data file 09   49,865 KB<br>Data file 10   54,825 KB |
| 7 | **Executable** (exe, bat, msi, etc.) | 1 KB to 100 MB | 10 | Executable 01   215 KB<br>Executable 02   3,597 KB<br>Executable 03   12,657 KB<br>Executable 04   19,417 KB<br>Executable 05   24,504 KB<br>Executable 06   37,745 KB<br>Executable 07   47,189 KB<br>Executable 08   75,022 KB<br>Executable 09   91,808 KB<br>Executable 10   94,732 KB |

### 3.2.3.2 Data Set 2

This group contains the following type of files, sizes, and quantities:

Table 3.2: Data Set 2

| Sr # | File Type | File Size | Quantity | |
|---|---|---|---|---|
| 1 | **Text files** (txt, doc, xls, pdf, etc.) | 100 MB to 2 GB | 5 | Text file 01 163,840 KB<br>Text file 02 470,016 KB<br>Text file 03 808,960 KB<br>Text file 04 1,126,400 KB<br>Text file 05 1,843,200 KB |
| 2 | **Media files** (mp3, mp4, m4v, etc.) | 100 MB to 2 GB | 5 | Media file 01 111,508 KB<br>Media file 02 238,816 KB<br>Media file 03 406,862 KB<br>Media file 04 760,782 KB<br>Media file 05 1,094,159 KB |
| 3 | **Archives** (zip, rar, 7z, etc.) | 100 MB to 2 GB | 5 | Archive 01 149,421 KB<br>Archive 02 257,068 KB<br>Archive 03 307,144 KB<br>Archive 04 603,964 KB<br>Archive 05 1,506,762 KB |
| 4 | **Disk images** (iso, bin, vcd, etc.) | 100 MB to 2 GB | 5 | Disk image 01 149,342 KB<br>Disk image 02 252,030 KB<br>Disk image 03 308,908 KB<br>Disk image 04 637,568 KB<br>Disk image 05 1,675,598 KB |
| 5 | **Data files** (dat, pst, sql, log, etc) | 100 MB to 2 GB | 5 | Data file 01 107,996 KB<br>Data file 02 186,052 KB<br>Data file 03 446,169 KB<br>Data file 04 751,443 KB<br>Data file 05 988,545 KB |

| 6 | **Executable** (exe, bat, msi, etc.) | 100 MB to 2 GB | 5 |  |

### 3.2.3.3 Data Set 3

This group contains the following type of files, sizes, and quantities:

Table 3.3: Data Set 3

| Sr # | File Type | File Size | Quantity | |
|---|---|---|---|---|
| 1 | **Text files** (txt, doc, xls, pdf, etc.) | 2 GB to 5 GB | 2 |  |
| 2 | **Media files** (mp3, mp4, m4v, etc.) | 2 GB to 5 GB | 2 |  |
| 3 | **Archives** (zip, rar, 7z, etc.) | 2 GB to 5 GB | 2 |  |
| 4 | **Disk images** (iso, bin, vcd, etc.) | 2 GB to 5 GB | 2 |  |
| 5 | **Data files** (dat, pst, sql, log, etc) | 2 GB to 5 GB | 2 |  |

### 3.2.3.4 Data Set 4

This group contains the following type of files, sizes, and quantities:

Table 3.4: Data Set 4

| Sr # | File Type | File Size | Quantity |
|---|---|---|---|

| 1 | **Text files** (txt, doc, xls, pdf, etc.) | Above 5 GB | 1 |  Text file 01  6,656,000 KB |
|---|---|---|---|---|
| 2 | **Media files** (mp3, mp4, m4v, etc.) | Above 5 GB | 1 |  Media file 01  6,767,922 KB |
| 3 | **Archives** (zip, rar, 7z, etc.) | Above 5 GB | 1 |  Archive 01  6,655,113 KB |
| 4 | **Disk images** (iso, bin, vcd, etc.) | Above 5 GB | 1 |  Disk image 01  6,936,576 KB |
| 5 | **Data files** (dat, pst, sql, log, etc) | Above 5 GB | 1 |  Data file 01  6,410,048 KB |

### 3.3 Work Flow

The workflow plays a significant role in any area of research. In this section, some basic requirements of the Operating System, tasks performed during research analysis are our main interest. A forensic workstation with commonly used Operating Systems like Microsoft Windows, Mac, or Linux is required. In the Microsoft Operating System, Windows 7 or onward version is compulsory, the reason being, TRIM only supported in Windows 7 or above versions. By default, the TRIM feature enabled in supported Operating Systems, so no configuration required. Tasks performed during the research are listed and explained in the figure below:

- Preparation of SSDs by using some professional disk wiping software
- Data storage and integrity check of all files, grouped in term of file types and sizes
- Data delete or drive format by using different available options
- Three image acquisitions after data deletion/format from a single SSD, using individual Data Set (1 out of 4) and with 1 minute, 1 hour, and 12 hours intervals
- Forensic analysis of drives and integrity verification in the case of data retrieval ▪ Result and conclusion

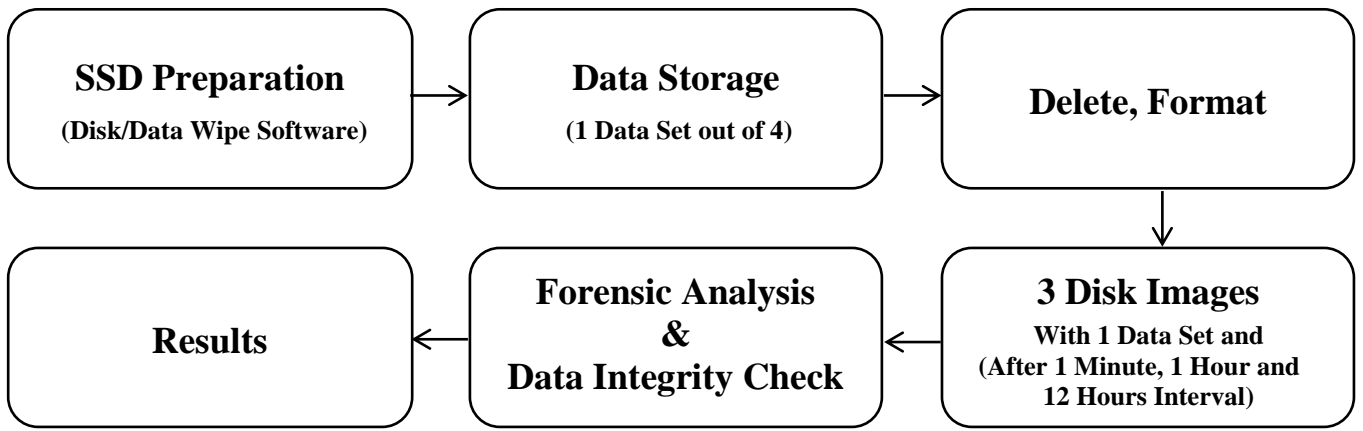

Figure 3.12: Work Flow

## 3.4 Summary

This section briefly covered the design of research work, methodologies, and the support of the TRIM feature in Operating Systems. Minimum requirements to carry out forensic analysis of SSDs and detailed specifications of hardware, software used also mentioned. In the last, the workflow diagram explained all phases and scenarios of overall work.

# Implementation and Results

## 4.1 Implementation

A systematic approach is mandatory to ease and harmonize all required tasks during scenarios of the entire research. The following are the steps used for forensic analysis of Solid State Drive (SSD) in different manufacturers of the same storage and technical specifications.

### 4.1.1 Preparation of Forensic Workstation

Clean installation of Microsoft Windows 7 Professional Edition, 64 bits, performed (optional but recommended). It removed unnecessary software, drivers, and third-party applications from a forensic workstation, which might affect forensic processes and results. Installation of only required software performed considered mandatory for the research. Installed caddy case in the laptop (forensic workstation) for mounting Solid State Drives (SSDs) and verified working TRIM command feature.

### 4.1.2 Preparation of Solid State Drive

Solid State Drive (SSD) needs to be wiped entirely by using a reliable disk wiping tool (with military-grade or maximum passes of erasing method). It is not mandatory if drives used the first time but recommended completely wipe so that all previous data on all drives deleted permanently. In this research work, all Solid State Drives (SSDs) thoroughly cleaned using *Erasure v 6.2* software with the Gutmann 35 passes method [18].

### 4.1.3 Data Storage in Solid State Drive

Disk wiping tool, i.e., Eraser, completely wiped previous data permanently from all Solid State Drives. At this moment, SSDs prepared for the storage of research data. Files of different types and sizes, already organized in the form of Data Sets copied to each disk for onward task execution. Data integrity of all files in each disk always checked by using *Checksum Compare v 1.42* software before the next task of drive format or data deletion.

### 4.1.4 Test Scenarios

Several test scenarios performed during forensic analysis of different Solid State Drives. All drives completely wiped with *Eraser* software before the copy of files from each *Data Set*s. Details of data deletion, the time between data deletion and image acquisition, tools used for forensic images and analysis, and format of disk image mentioned below:

- **Data Delete**

    In each test scenario, data deleted with the following available options:

    - Delete
    - Shift + Delete

- **Drive Format**

    In each test scenario, drive formatted with the following available options:

    - Format with Windows Explorer built-in option
    - Format with SSD manufacturer's software utility

- **Time Interval**

    There is no general rule and requirement to fix specific time between data deletion and image acquisition, but for the research and analysis of Wear Leveling in different Solid State Drives, *1 minute, 1 hour, and 12 hours* time interval selected. During this time, a forensic workstation (laptop in this case) remained in a power-on state and with no other activity performed.

- **Imaging Tool and Format**

    - In each test scenario, *after data deletion*, images of all Solid State Drives (SSDs) acquired after 1 minute, 1 hour, and 12 hours time interval with *FTK Imager v 4.2* software and in *Expert Witness (.e01)* format.

    - In each test scenario, *after the format of drives*, images of all Solid State Drives (SSDs) acquired after only 1 minute of the time interval with *FTK Imager v 4.2* software and in *Expert Witness (.e01)* format.

## 4.1.5 Disk Images

Following is the summary of disk images acquired during throughout research work:

- **Images of Single Disk**
  - **TRIM:** On (default)
  - **Disk:** SSD 1 (out of 3)
  - **Data:** Data Set 1 (out of 4)
  - **Mode:** Delete / Format
  - **Time interval:** 1 minute, 1 hour and 12 hours

| | |
|---|---|
| Disk images (1 Data Set, Delete): | 3 |
| Disk images (4 Data Sets, Delete): | 4 x 3 =  12 |
| Disk images (4 Data Sets, Format, 1 Min): | 4 |
| **Total images of single SSD:** | 4 + 12 =  16 |
| **Total images of all SSDs:** | 16 x 3 =  48 |

Table 4.1: Test Scenarios

| Scenarios | Interval | Mode | SSD 1 | SSD 2 | SSD 3 |
|---|---|---|---|---|---|
| **Data Set 1**<br>1 KB to 100 MB | 1 Minute | Delete | √ | √ | √ |
| | 1 Hour | Delete | √ | √ | √ |
| | 12 Hours | Delete | √ | √ | √ |
| | 1 Minute | Format | √ | √ | √ |
| **Data Set 2**<br>100 MB to 2 GB | 1 Minute | Delete | √ | √ | √ |
| | 1 Hour | Delete | √ | √ | √ |
| | 12 Hours | Delete | √ | √ | √ |
| | 1 Minute | Format | √ | √ | √ |
| **Data Set 3**<br>2 GB to 5 GB | 1 Minute | Delete | √ | √ | √ |
| | 1 Hour | Delete | √ | √ | √ |
| | 12 Hours | Delete | √ | √ | √ |
| | 1 Minute | Format | √ | √ | √ |
| **Data Set 4**<br>Above 5 GB | 1 Minute | Delete | √ | √ | √ |
| | 1 Hour | Delete | √ | √ | √ |
| | 12 Hours | Delete | √ | √ | √ |
| | 1 Minute | Format | √ | √ | √ |

## 4.2  Results

To analyze the behavior of Wear Leveling in manufactures of different Solid State Drives (SSDs) with the same technical specifications, files of different types, sizes, and time of disk image acquisition are the main concerns of this research. By default, the TRIM command feature in the Operating System of the forensic workstation is on so that in case of data delete or drive format, message to SSD controller will deliver informing about the erasure of data from drives cells. The garbage collection process will execute hereafter to release all unused pages for future utilization. Following is the results of all research scenarios conducted in manufacturers of different Solid State Drives (SSDs):

### 4.2.1  Solid State Drive (SSD) 1
(Micron P400e, 2.5" 64 GB)

#### 4.2.1.1   Micron SSD with Data Set 1

In this scenario, different types of files and their sizes from **1 KB to 100 MB** used. All files of *Data Set 1* saved in the drive and verified data integrity. *Deleted* all the files and acquired disk images with *FTK Imager* software after 1 minute, 1 hour, and 12 hours duration. Analyzed disk images and their results in graphical representation provided below:

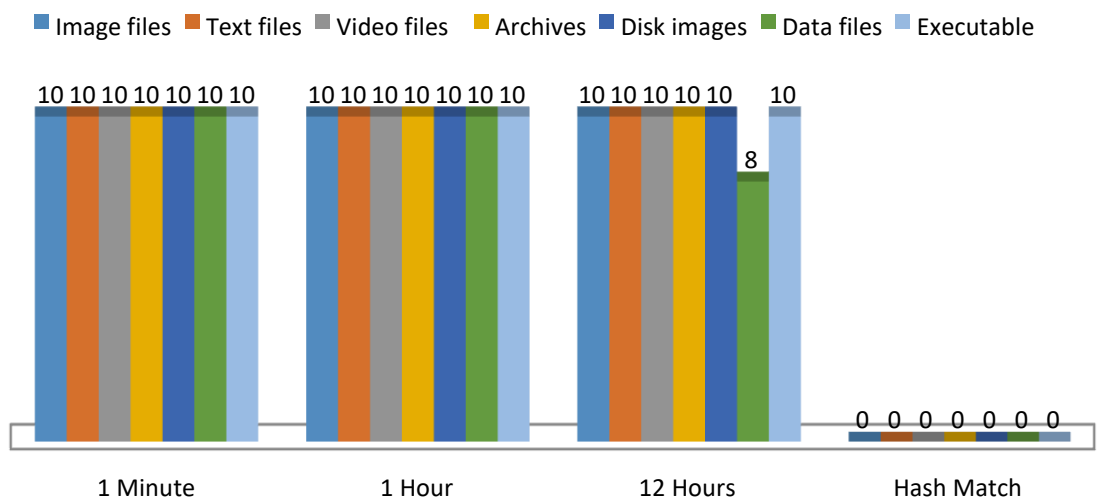

Figure 4.1:  Micron SSD with Data Set 1

**Analysis:** From *Data Set 1*, total numbers of seventy (70) files saved in the drive. After data delete and forensic analysis of all three images, not a single file with the same original file hash recovered. The process of Wear Leveling and Garbage Collection started within the first minute; as a result, all files recovered in corrupt form. In a couple of hours, two (02) **database** files wiped out completely and left no residual information in the drive.

### 4.2.1.2  Micron SSD with Data Set 2

In this scenario, different types of files and their sizes from **100 MB to 2 GB** used. Drive was fully erased with *Erasure* software before any files storage so that previous data in the drive wiped out completely. All files of *Data Set 2* saved in the drive and verified data integrity. *Deleted* all the files and acquired disk images with *FTK Imager* software after 1 minute, 1 hour, and 12 hours duration. Analyzed disk images and their results in graphical representation provided below:

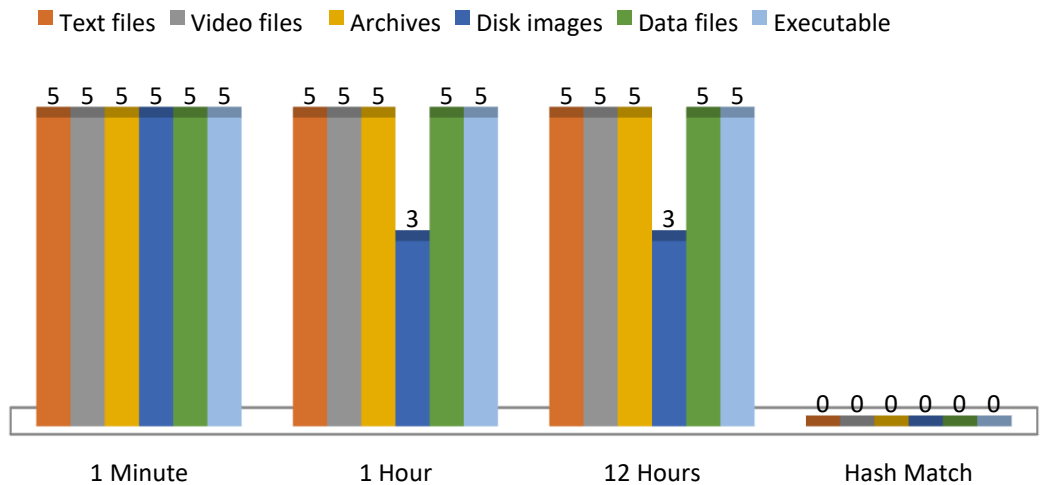

Figure 4.2:  Micron SSD with Data Set 2

**Analysis:** From *Data Set 2*, total numbers of thirty (30) files saved in the drive. After data delete and forensic analysis of all three images, not a single file with the same original file hash recovered. The process of Wear Leveling and Garbage Collection started within the first minute; as a result, all files recovered in corrupt form. In the first hour after data deletion, two
(02) **disk image** files wiped out completely and left no residual information in the drive.

### 4.2.1.3 Micron SSD with Data Set 3

In this scenario, different types of files and their sizes from **2 GB to 5 GB** used. Drive was fully erased with *Erasure* software before any files storage so that previous data in the drive wiped out completely. All files of *Data Set 3* saved in the drive and verified data integrity. *Deleted* all the files and acquired disk images with *FTK Imager* software after 1 minute, 1 hour, and 12 hours duration. Analyzed disk images and their results in graphical representation provided below:

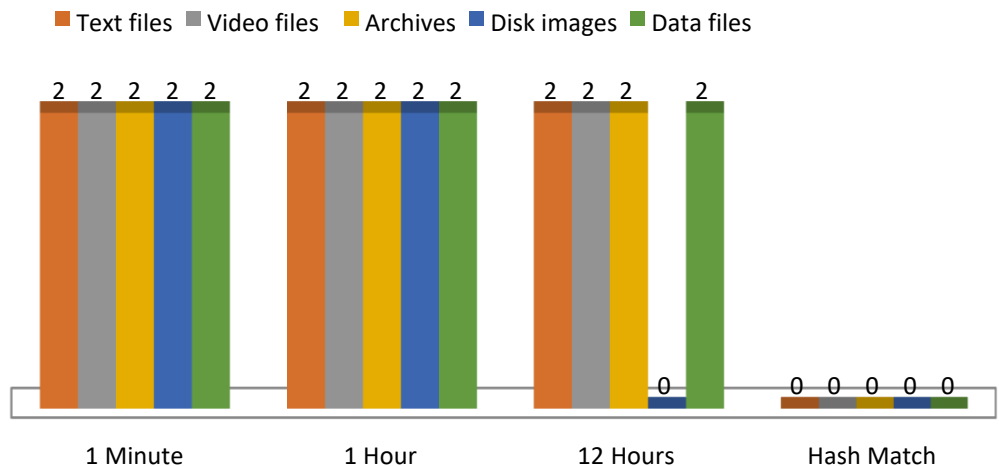

Figure 4.3: Micron SSD with Data Set 3

**Analysis:** From *Data Set 3*, total numbers of ten (10) files saved in the drive. After data delete and forensic analysis of all three images, not a single file with the same original file hash recovered. The process of Wear Leveling and Garbage Collection started within the first minute; as a result, all files recovered in corrupt form. In a couple of hours, two (02) **disk image** files wiped out completely and left no residual information in the drive.

### 4.2.1.4 Micron SSD with Data Set 4

In this scenario, different types of files and their sizes **above 5 GB** used. Drive was fully erased with *Erasure* software before any files storage so that previous data in the drive wiped out completely. All files of *Data Set 4* saved in the drive and verified data integrity. *Deleted* all the files and acquired disk images with *FTK Imager* software after 1 minute, 1 hour, and 12 hours duration. Analyzed disk images and their results in graphical representation provided below:

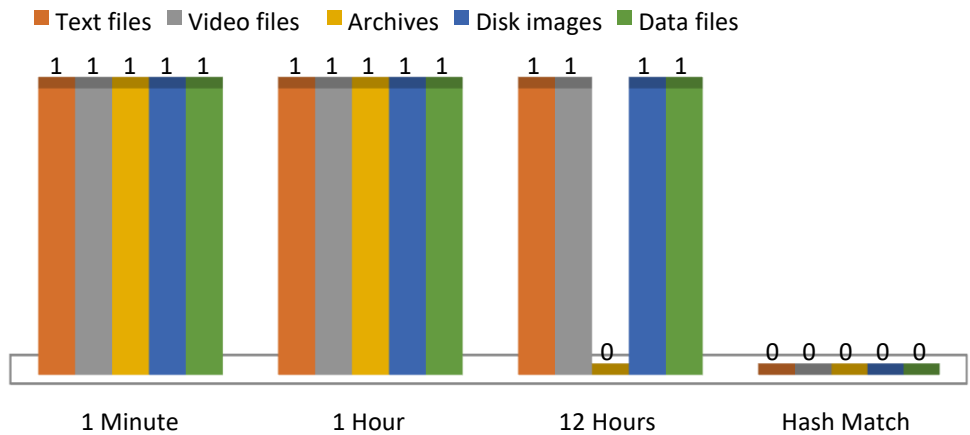

Figure 4.4: Micron SSD with Data Set 4

**Analysis:** From *Data Set 4*, total numbers of five (05) files saved in the drive. After data delete and forensic analysis of all three images, not a single file with the same original file hash recovered. The process of Wear Leveling and Garbage Collection started within the first minute; as a result, all files recovered in corrupt form. In a couple of hours, one (01) **archive** file wiped out completely and left no residual information in the drive.

### 4.2.1.5  Research Summary of Micron SSD

In SSD 1 (Micron), files of different types and sizes used to analyze the forensic behavior of Wear Leveling, TRIM, and Garbage Collection processes. In the test scenario with disk **format** option, not a single file with any information retrieved from the drive.

In all test scenarios of Micron SSD with data **delete** options, entire data with original file sizes was available to recover, but all of the files were in a corrupt state. Garbage Collection process executed immediately after data delete, and its behavior on files was different in terms of their type and sizes.

In Data Set 1 (1 KB – 100 MB), database files behaved differently than others and wiped completely from the drive within a couple of hours. In Data Set 2 (100 MB – 2 GB) disk image files wiped completely within the first hour. It was the same in the case of Data Set 3 (2 GB – 5 GB) files with some more time duration, and in the above 5 GB files group (Data Set 4), the behavior of archive files vary and wipe out after some hours from the drive.

### 4.2.2    Solid State Drive (SSD) 2
(SanDisk SDSSDP, 2.5" 64 GB)

#### 4.2.2.1    SanDisk SSD with Data Set 1

In this scenario, different types of files and their sizes from **1 KB to 100 MB** used. All files of *Data Set 1* saved in the drive and verified data integrity. *Deleted* all the files and acquired disk images with *FTK Imager* software after 1 minute, 1 hour, and 12 hours duration. Analyzed disk images and their results in graphical representation provided below:

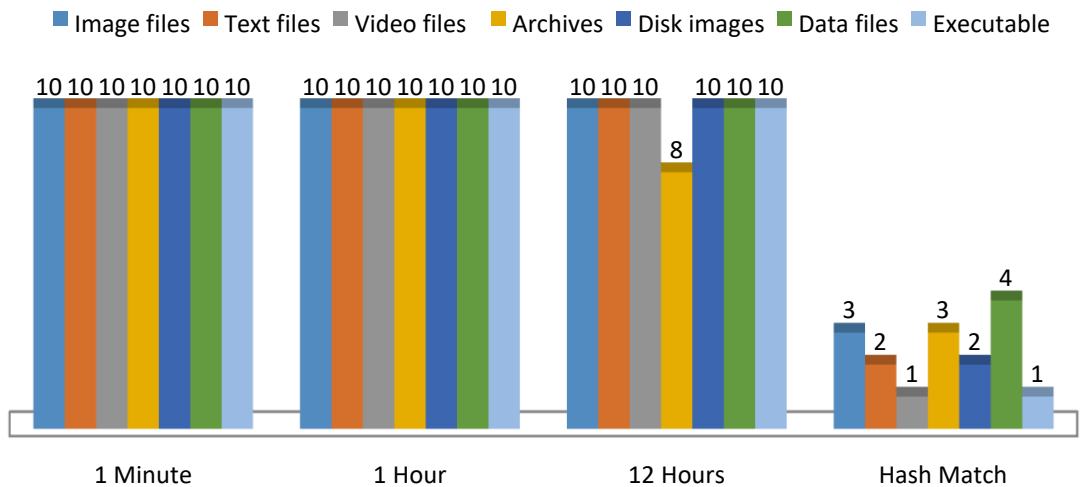

Figure 4.5: SanDisk SSD with Data Set 1

**Analysis:** From *Data Set 1*, total numbers of seventy (70) files saved in the drive. After data delete and forensic analysis of all three images, sixteen (16) files with the same original files hashes recovered successfully. Wear Leveling and Garbage Collection processes started within the first minute; as a result, corrupted some files of all available types. In a couple of hours, two (02) **archive** files wiped out completely and left no residual information in the drive.

#### 4.2.2.2    SanDisk SSD with Data Set 2

In this scenario, different types of files and their sizes from **100 MB to 2 GB** used. Drive was fully erased with *Erasure* software before any files storage so that previous data in the drive wiped out completely. All files of *Data Set 2* saved in the drive and verified data integrity.

*Deleted* all the files and acquired disk images with *FTK Imager* software after 1 minute, 1 hour, and 12 hours duration. Analyzed disk images and their results in graphical representation provided below:

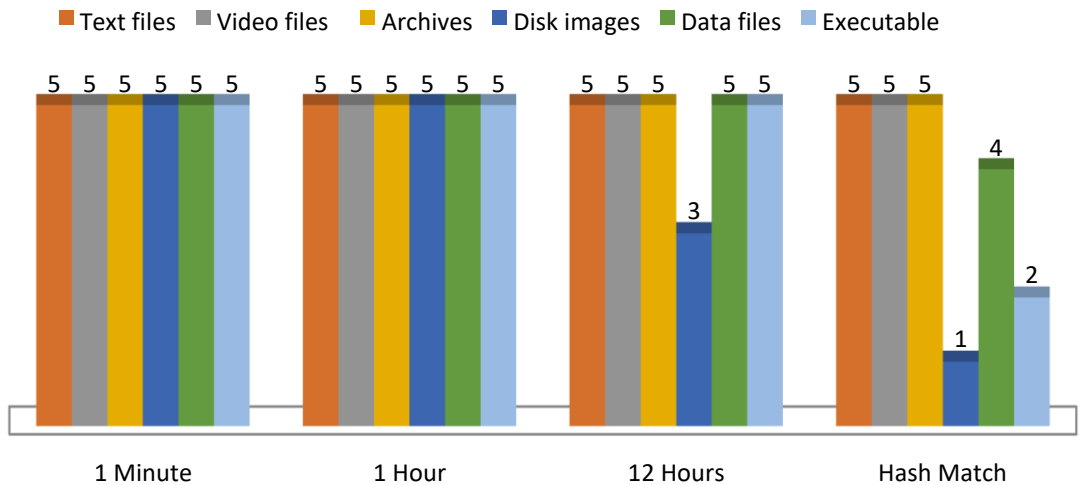

Figure 4.6: SanDisk SSD with Data Set 2

**Analysis:** From *Data Set 2*, total numbers of thirty (30) files saved in the drive. After data delete and forensic analysis of all three images, twenty-two (22) files with the same original files hashes recovered successfully. Wear Leveling and Garbage Collection processes started within the first minute; as a result, corrupted some disk image, database, and executable files only. In a couple of hours, two (02) **disk image** files wiped out completely and left no residual information in the drive.

### 4.2.2.3 SanDisk SSD with Data Set 3

In this scenario, different types of files and their sizes from **2 GB to 5 GB** used. Drive was fully erased with *Erasure* software before any files storage so that previous data in the drive wiped out completely. All files of *Data Set 3* saved in the drive and verified data integrity. *Deleted* all the files and acquired disk images with *FTK Imager* software after 1 minute, 1 hour, and 12 hours duration. Analyzed disk images and their results in graphical representation provided below:

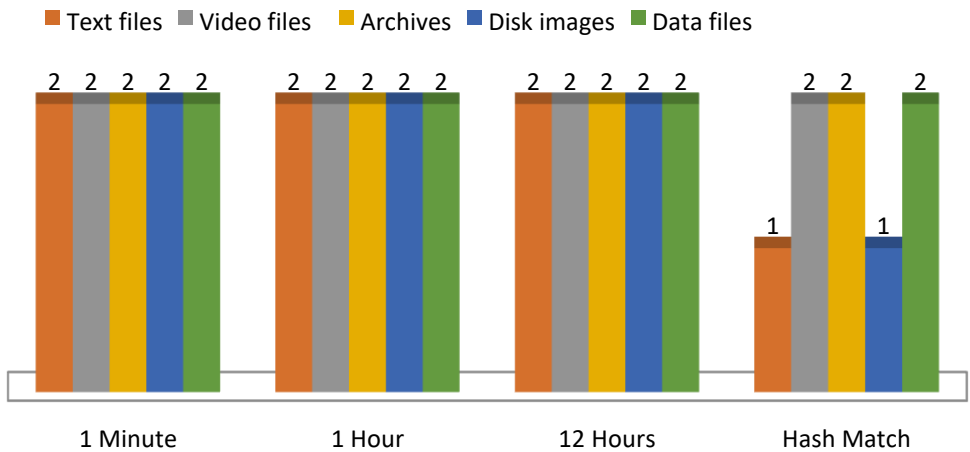

Figure 4.7: SanDisk SSD with Data Set 3

**Analysis:** From *Data Set 3*, total numbers of ten (10) files saved in the drive. After data delete and forensic analysis of all three images, eight (08) files with the same original files hashes recovered successfully. Wear Leveling and Garbage Collection processes started within the first minute; as a result, corrupted some text and disk image files only. No file completely wiped out in 12 hours duration.

### 4.2.2.4 SanDisk SSD with Data Set 4

In this scenario, different types of files and their sizes **above 5 GB** used. Drive was fully erased with *Erasure* software before any files storage so that previous data in the drive wiped out completely. All files of *Data Set 4* saved in the drive and verified data integrity. *Deleted* all the files and acquired disk images with *FTK Imager* software after 1 minute, 1 hour, and 12 hours duration. Analyzed disk images and their results in graphical representation provided below:

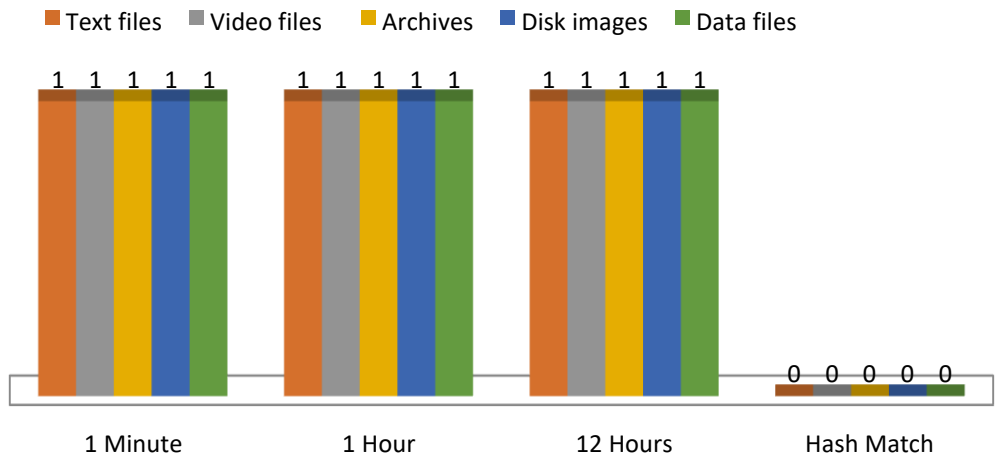

Figure 4.8: SanDisk SSD with Data Set 4

**Analysis:** From *Data Set 4*, total numbers of five (05) files saved in the drive. After data delete and forensic analysis of all three images, not a single file with the same original file hash recovered. Wear Leveling and Garbage Collection processes started within the first minute; as a result, corrupted all the files. No file completely wiped out in 12 hours duration.

#### 4.2.2.5 Research Summary of SanDisk SSD

In SSD 2 (SanDisk), files of the same data sets used to analyze the forensic behavior of Wear Leveling, TRIM, and Garbage Collection processes. In the test scenario with disk **format** option, not a single file with any information retrieved from the drive.

In all test scenarios of SanDisk SSD with data **delete** options, the majority of data was available to restore completely with their same original files hashes. Some types of files recovered in a corrupt state, and very few files wiped out completely from the drive. Garbage Collection process executed after a lapse of time in this case, and its behavior on the files was different in terms of their type and sizes.

In Data Set 1 (1 KB – 100 MB), archive files wiped completely from the drive within a couple of hours; similarly, in Data Set 2 (100 MB – 2 GB), disk image files wiped completely after some hours. In Data Set 3 and Data Set 4, none of any files wiped out within a 12-hour time duration.

### 4.2.3 Solid State Drive (SSD) 3

(Transcend SSD370S, 2.5" 64 GB)

#### 4.2.3.1 Transcend SSD with Data Set 1

In this scenario, different types of files and their sizes from **1 KB to 100 MB** used. All files of *Data Set 1* saved in the drive and verified data integrity. *Deleted* all the files and acquired disk images with *FTK Imager* software after 1 minute, 1 hour, and 12 hours duration. Analyzed disk images and their results in graphical representation provided below:

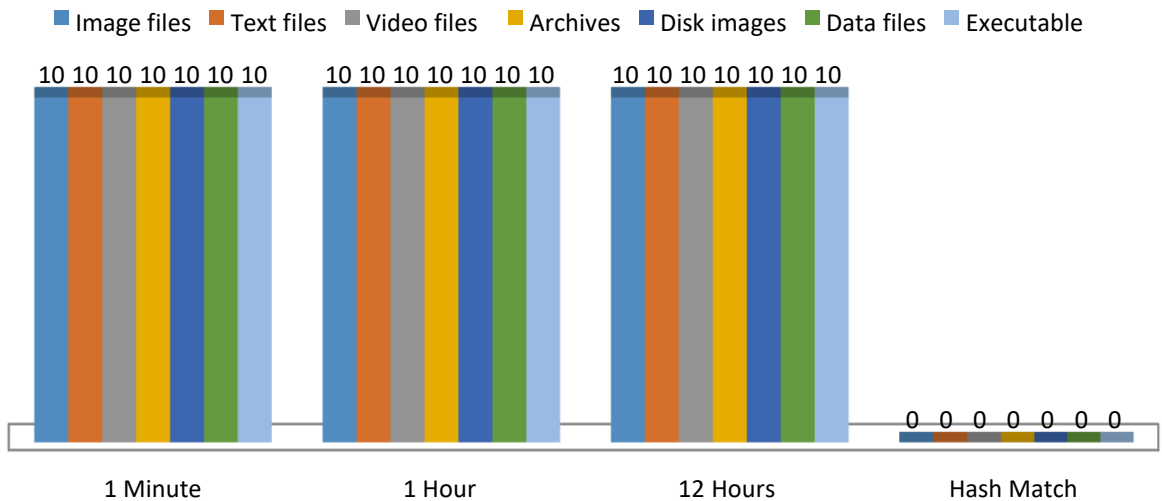

Figure 4.9: Transcend SSD with Data Set 1

**Analysis:** From *Data Set 1*, total numbers of seventy (70) files saved in the drive. After data delete and forensic analysis of all three images, not a single file with the same original file hash recovered. The process of Wear Leveling and Garbage Collection started within the first minute; as a result, all files recovered in corrupt form. No file completely wiped out in 12 hours duration.

#### 4.2.3.2 Transcend SSD with Data Set 2

In this scenario, different types of files and their sizes from **100 MB to 2 GB** used. Drive was fully erased with *Erasure* software before any files storage so that previous data in the drive wiped out completely. All files of *Data Set 2* saved in the drive and verified data integrity. *Deleted* all the files and acquired disk images with *FTK Imager* software after 1 minute, 1 hour, and 12 hours duration. Analyzed disk images and their results in graphical representation provided below:

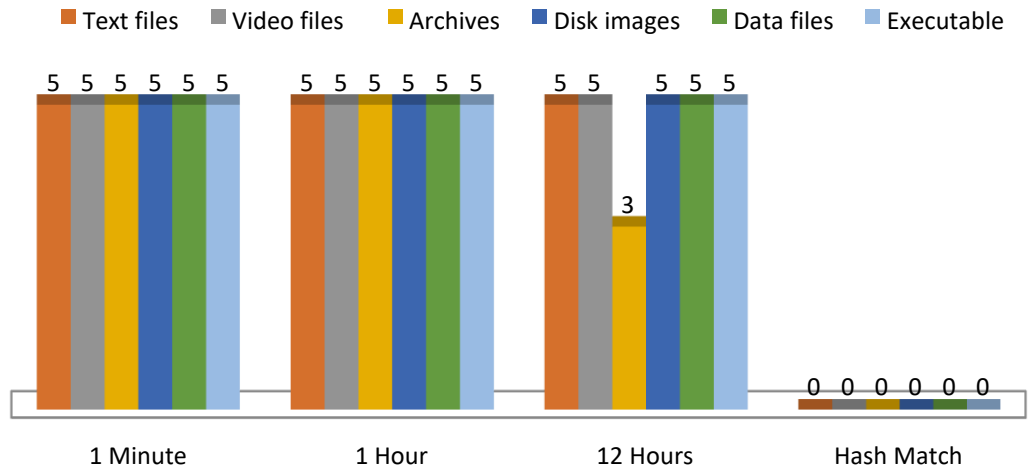

Figure 4.10: Transcend SSD with Data Set 2

**Analysis:** From *Data Set 2*, total numbers of thirty (30) files saved in the drive. After data delete and forensic analysis of all three images, not a single file with the same original file hash recovered. The process of Wear Leveling and Garbage Collection started within the first minute; as a result, all files recovered in corrupt form. In a couple of hours, two (02) **archive** files wiped out completely and left no residual information in the drive.

### 4.2.3.3  Transcend SSD with Data Set 3

In this scenario, different types of files and their sizes from **2 GB to 5 GB** used. Drive was fully erased with *Erasure* software before any files storage so that previous data in the drive wiped out completely. All files of *Data Set 3* saved in the drive and verified data integrity. *Deleted* all the files and acquired disk images with *FTK Imager* software after 1 minute, 1 hour, and 12 hours duration. Analyzed disk images and their results in graphical representation provided below:

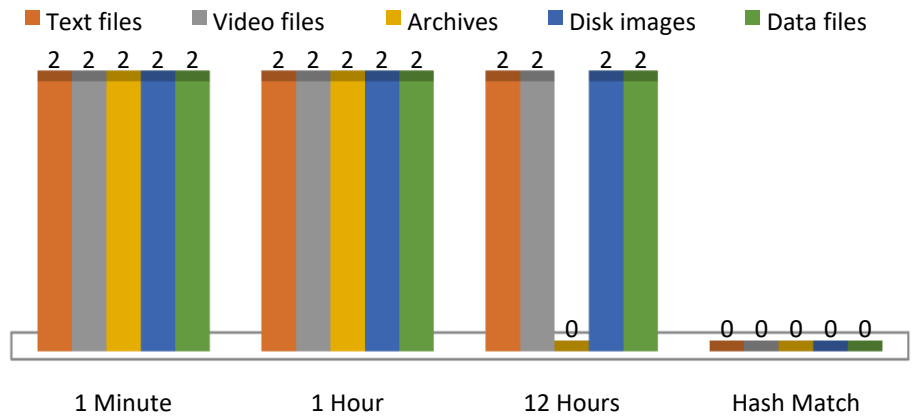

Figure 4.11: Transcend SSD with Data Set 3

**Analysis:** From *Data Set 3*, total numbers of ten (10) files saved in the drive. After data delete and forensic analysis of all three images, not a single file with the same original file hash recovered. The process of Wear Leveling and Garbage Collection started within the first minute; as a result, all files recovered in corrupt form. In a couple of hours, two (02) **archive** files wiped out completely and left no residual information in the drive.

#### 4.2.3.4 Transcend SSD with Data Set 4

In this scenario, different types of files and their sizes **above 5 GB** used. Drive was fully erased with *Erasure* software before any files storage so that previous data in the drive wiped out completely. All files of *Data Set 4* saved in the drive and verified data integrity. *Deleted* all the files and acquired disk images with *FTK Imager* software after 1 minute, 1 hour, and 12 hours duration. Analyzed disk images and their results in graphical representation provided below:

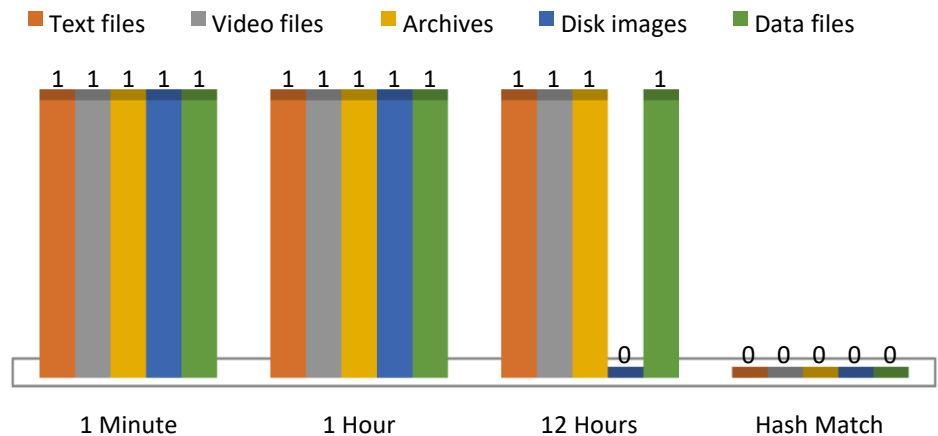

Figure 4.12: Transcend SSD with Data Set 4

**Analysis:** From *Data Set 4*, total numbers of five (05) files saved in the drive. After data delete and forensic analysis of all three images, not a single file with the same original file hash recovered. The process of Wear Leveling and Garbage Collection started within the first minute; as a result, all files recovered in corrupt form. In a couple of hours, one (01) **disk image** file wiped out completely and left no residual information in the drive.

### 4.2.3.5  Research Summary of Transcend SSD

In SSD 3 (Transcend), files of the same data sets used to analyze the forensic behavior of Wear Leveling, TRIM, and Garbage Collection processes. In the test scenario with disk **format** option, not a single file with any information retrieved from the drive.

In all test scenarios of Transcend SSD with data **delete** options, entire data with original file sizes was available to restore, but all of the files were in a corrupt state. Garbage Collection process executed immediately after data delete, and its behavior on files was different in terms of their type and sizes.

In Data Set 1 (1 KB – 100 MB), no file wiped from the drive within a 12-hour time duration. In Data Set 2 and Data Set 3, archive files wiped completely from the drive within a couple of hours. In Data Set 4 (Above 5 GB) disk image files behaved differently and wiped out completely after a couple of hours.

### 4.2.4 Results Overview

Following table provides details of the result acquired from all Solid State Drives:

Table 4.2: Result of All SSDs

| Scenarios | Description | SSD 1 (Micron) | SSD 2 (SanDisk) | SSD 3 (Transcend) |
|---|---|---|---|---|
| Data Set 1 Delete (1 Min) | 1 KB to 100 MB<br><br>File types: 7 (Image, Text, Video, Archive, Disk Image, Data, and Executable)<br><br>Total files: 7 x 10 = 70 | Files remain: All (70)<br>Files sizes: Same<br>Health: Corrupted<br>Hash: All Mismatch | Files remain: All (70)<br>Files sizes: Same<br>Health: 23/70 Good<br>Hash: 23 Match | Files remain: All (70)<br>Files sizes: Same<br>Health: Corrupted<br>Hash: All Mismatch |
| Data Set 1 Delete (1 Hour) | | Files remain: All (70)<br>Files sizes: Same<br>Health: Corrupted<br>Hash: All Mismatch | Files remain: All (70)<br>Files sizes: Same<br>Health: 17/70 Good<br>Hash: 17 Match | Files remain: All (70)<br>Files sizes: Same<br>Health: Corrupted<br>Hash: All Mismatch |
| Data Set 1 Delete (12 Hours) | | Files remain: 68/70 (2 lost)<br>Files sizes: Same<br>Health: Corrupted<br>Hash: All Mismatch | Files remain: 68/70 (2 lost)<br>Files sizes: Same<br>Health: 16/70 Good<br>Hash: 16 Match | Files remain: All (70)<br>Files sizes: Same<br>Health: Corrupted<br>Hash: All Mismatch |
| Data Set 1 Format (1 Minute) | | Files remain: NIL | Files remain: NIL | Files remain: NIL |
| Data Set 2 Delete (1 Minute) | 100 MB to 2 GB<br><br>File types: 6 (Text, Video, Archive, Disk Image, Data, and Executable)<br><br>Total files: 6 x 5 = 30 | Files remain: All (30)<br>Files sizes: Same<br>Health: Corrupted<br>Hash: All Mismatch | Files remain: All (30)<br>Files sizes: Same<br>Health: 22/30 Good<br>Hash: 22 Match | Files remain: All (30)<br>Files sizes: Same<br>Health: Corrupted<br>Hash: All Mismatch |
| Data Set 2 Delete (1 Hour) | | Files remain: 28/30 (2 lost)<br>Files sizes: Same<br>Health: Corrupted<br>Hash: All Mismatch | Files remain: All (30)<br>Files sizes: Same<br>Health: 22/30 Good<br>Hash: 22 Match | Files remain: All (30)<br>Files sizes: Same<br>Health: Corrupted<br>Hash: All Mismatch |
| Data Set 2 Delete (12 Hours) | | Files remain: 28/30 (2 lost)<br>File sizes: Same<br>Health: Corrupted<br>Hash: All Mismatch | Files remain: 28/30 (2 lost)<br>Files sizes: Same<br>Health: 22/30 Good<br>Hash: 22 Match | Files remain: 28/30 (2 lost)<br>Files sizes: Same<br>Health: Corrupted<br>Hash: All Mismatch |
| **Data Set 2 Format (1 Minute)** | | **Files remain:** NIL | **Files remain:** NIL | **Files remain:** NIL |
| **Data Set 3 Delete (1 Minute)** | **2 GB to 5 GB** **File types:** 5 (Text, Video, Archive, Disk Image, and Data) **Total files:** 5 x 2 = 10 | **Files remain:** All (10) **Files sizes:** Same **Health:** Corrupted **Hash:** All Mismatch | **Files remain:** All (10) **Files sizes:** 1/10 differ **Health:** 8/10 Good **Hash:** 8 Match | **Files remain:** All (10) **Files sizes:** 1/10 differ **Health:** Corrupted **Hash:** All Mismatch |
| **Data Set 3 Delete (1 Hour)** | | **Files remain:** All (10) **Files sizes:** Same **Health:** Corrupted **Hash:** All Mismatch | **Files remain:** All (10) **Files sizes:** 1/10 differ **Health:** 8/10 Good **Hash:** 8 Match | **Files remain:** All (10) **Files sizes:** 1/10 differ **Health:** Corrupted **Hash:** All Mismatch |
| **Data Set 3 Delete (12 Hours)** | | **Files remain:** 8/10 (2 lost) **Files sizes:** 1/10 differ **Health:** Corrupted **Hash:** All Mismatch | **Files remain:** All (10) **Files sizes:** 1/10 differ **Health:** 8/10 Good **Hash:** 8 Match | **Files remain:** 8/10 (2 lost) **Files sizes:** 1/10 differ **Health:** Corrupted **Hash:** All Mismatch |
| **Data Set 3 Format (1 Minute)** | | **Files remain:** NIL | **Files remain:** NIL | **Files remain:** NIL |
| **Data Set 4 Delete (1 Minute)** | **Above 5 GB** **File types:** 5 (Text, Video, Archive, Disk Image, and Data) **Total files:** 5 x 1 = 5 | **Files remain:** All (5) **Files sizes:** 5/5 differ **Health:** Corrupted **Hash:** All Mismatch | **Files remain:** All (5) **Files sizes:** 5/5 differ **Health:** Corrupted **Hash:** All Mismatch | **Files remain:** All (5) **Files sizes:** 5/5 differ **Health:** Corrupted **Hash:** All Mismatch |
| **Data Set 4 Delete (1 Hour)** | | **Files remain:** All (5) **Files sizes:** 5/5 differ **Health:** Corrupted **Hash:** All Mismatch | **Files remain:** All (5) **Files sizes:** 5/5 differ **Health:** Corrupted **Hash:** All Mismatch | **Files remain:** All (5) **Files sizes:** 5/5 differ **Health:** Corrupted **Hash:** All Mismatch |
| **Data Set 4 Delete (12 Hours)** | | **Files remain:** 4/5 (1 lost) **Files sizes:** 5/5 differ **Health:** Corrupted **Hash:** All Mismatch | **Files remain:** All (5) **Files sizes:** 5/5 differ **Health:** Corrupted **Hash:** All Mismatch | **Files remain:** 4/5 (1 lost) **Files sizes:** 5/5 differ **Health:** Corrupted **Hash:** All Mismatch |

| Data Set 4 Format (1 Minute) | Files remain: NIL | Files remain: NIL | Files remain: NIL |

# Conclusion and Future Work

## 5.1 Conclusion

In consideration of results obtained from the experiments, it concluded that the behavior of Wear Leveling in different SSD manufacturers having the same storage capacities does not match. It varies based on the number of files, types of files, and sizes. The recovery of files from different SSD manufacturers showed different results. In all SSDs, not a single trace of any file found in disk format scenario(s). Whereas, some of the data recovered in the delete case and from only one drive. It clearly showed different behavior of data recoveries in format and delete cases.

The obvious finding from this study is that the time interval of image acquisitions played a significant role, and the longer time interval supports few chances of data recovery because the TRIM and Garbage Collection process effects clearing residual data from the drives.

Reference to our work scope and to find the answers to our research questions, the following is the conclusion of this research:

***Research Question 1:*** *Is the behavior of Wear Leveling in different SSD manufacturers having the same drive capacities equivalent?*

The behavior of Wear Leveling is not the same in different SSDs, even with exact technical specifications like identical flash chips and same storage capacities. Two of the SSDs (Micron and Transcend) showed almost the same behavior, but the third one (SanDisk) was entirely different. It leads to the conclusion that the behavior of Wear Leveling in different SSD manufacturers having the same drive capacities is not equivalent.

***Research Question 2:*** *Does the file size and type affect the evidence recovery in SSDs from different manufacturers?*

Different file types and sizes from 1 KB to above 5 GB, selected for the research and grouped them into four Data Sets. Analyzed the behavior of these files in different drives and concluded that the file types and their sizes clearly affect in evidence recovery from manufacturers of different Solid State Drives even with the same storage capacities.

***Research Question 3:*** *Do the formatting and data deletion in different SSDs affect the evidence recovery?*

Delete and format are two methods of erasing data in a directory or from the drive. Data delete refers to the only removal of the file(s) individually, but drive format leads to overall erasure of disk or partition at once. In this research, both options were used and found different results. If drives format with quick or full mode, or even with vendor-specific tools, not a single file or any information recovered from the drives. In the data delete option, some of the files retrieved successfully along with originally matched file hashes. Therefore, the conclusion is that both options behaved differently and makes an affect in evidence recovery.

***Research Question 4:*** *Is the time difference between the data deletion and image acquisition affect the evidence recovery process outcome?*

In this research, the time of drives image acquisition after data delete or drive format selected as 1 minute, 1 hour, and 12 hours duration. In these different times, the behavior of all three drives was not the same. Micron and Transcend SSD's response was almost identical, but SanDisk's behavior was entirely different. In this drive, the Garbage Collection process executed after some hours, but in other drives, the same process performed within the first minute and destroyed all the data. Therefore it is concluded that time delay after data delete and image acquisition plays a significant role; it varies in different drives and affects the evidence recovery process outcome.

## 5.2   Future Work

In the future, the same research can extend by using high capacity industrial grade Solid State Drives (SSD) with different Operating Systems like Windows 10, Linux, and Mac OS. A time interval of data delete, and disk images acquisition can be more than 12 hours to observe the forensic behavior of time factor among different SSD manufacturers. Besides, the NTFS file system on the drives, research can expand to modern file systems. More research options are available by using proprietary forensic tools like SANS SIFT, Forensic Took Kit (FTK), and EnCase Forensic software.